\documentclass[11pt]{article}
   
\usepackage{epsfig}
\usepackage{amsfonts,amssymb}

\renewcommand{\theequation}{\arabic{section}.\arabic{equation}}

\font\mybb=msbm10 at 12pt 
\def\bb#1{\hbox{\mybb#1}}

\newcommand{\del}{\ensuremath{\partial}}

\newcommand{\half}{\ensuremath{\frac{1}{2}}}

\newcommand{\quarter}{\ensuremath{\frac{1}{4}}}

\def\PM{{\mathbb P}}
\def\CM{{\mathbb C}}

\newcommand{\be}{\begin{equation}}
\newcommand{\ee}{\end{equation}}

\newcommand{\ba}{\begin{eqnarray}}
\newcommand{\ea}{\end{eqnarray}}

\newcommand{\ns}{\normalsize}

\begin{document}


\begin{titlepage}

\title{
   \hfill{\ns hep-th/0411033\\}
   \vskip 2cm
   {\Large\bf Type IIB Conifold Transitions in Cosmology}
\\[0.5cm]}
   \setcounter{footnote}{0}
\author{
{\ns\large 
  \setcounter{footnote}{3}
  Andre Lukas$^1$\footnote{email: lukas@physics.ox.ac.uk}~, Eran Palti$^2$\footnote{email: e.palti@sussex.ac.uk}~, P.M.Saffin$^2$\footnote{email: p.m.saffin@sussex.ac.uk}}
\\[0.5cm]
   {$^1$\it\ns Rudolf Peierls Centre for Theoretical Physics, University
    of Oxford}\\
   {\ns 1 Keble Road, Oxford OX1 3NP, UK}\\[0.3cm]
   {$^2$\it\ns Department of Physics and Astronomy, University of Sussex}\\
   {\ns Falmer, Brighton BN1 9QJ, UK} \\[0.2em] }
\date{}

\maketitle

\begin{abstract}\noindent

We study four-dimensional low energy effective actions for conifold
transitions of Calabi-Yau spaces in the context of IIB supergravity.
The actions are constucted by examining the mass of D3-branes wrapped
on collapsing/expanding three-cycles. We then study the cosmology
of the conifold transition, including consequences for moduli stabilization,
taking into account the effect of the additional states which become
light at the transition. We find, the degree to which the additional
states are excited is essential for whether the transition is 
dynamically realized.
\end{abstract}

\thispagestyle{empty}

\end{titlepage}

\section{Introduction}

The problem of string theory vacuum degeneracy splits into two parts,
namely the continuous degeneracy due to moduli fields and the discrete
one due to the large number of different topologies. As is well-known
by now there are a number of different topology-changing processes
\cite{Candelas:1988di}\cite{Candelas:1989ug}\cite{Greene:1996cy} in string theory which connect the moduli spaces
associated with topologically different compactifications. A detailed
understanding of these processes may perhaps be considered as a first
step towards resolving the topological degeneracy. In this paper, we
will study a certain type of topology-changing process, the conifold
transition, as an explicitly time-dependent phenomena in the context
of string cosmology. 

A ``milder'' type of topology changing transition which arises in the
context of type II Calabi-Yau compactifications is referred to as
``flop''.  In such a transition, a two-cycle within the Calabi-Yau
space contracts to a point and re-expands as another two-cycle,
thereby leading to a topologically distinct Calabi-Yau space with
different intersection numbers. However, on either side of the
transition the Hodge numbers are the same and so the spectrum of
massless moduli is unchanged, while only their interactions are
affected.  The study of 5D cosmology associated with a flop of the
internal space in M-theory was initiated in \cite{Brandle:2002fa} and
followed up in \cite{Jarv:2003qx}\cite{Jarv:2003qy}. There it was
found that the tension of any M2-branes wrapping the collapsing
two-cycles caused the cycles to remain small, so the moduli forced the
CY to remain near the flop point.  Earlier work on black holes where
the internal manifold undergoes a flop as a function of radius is
given in \cite{Gaida:1998km}.

In the conifold transition, which we study here, a three-cycle on one
side of the transition collapses and then re-expands into a two-cycle,
thus changing the Hodge numbers of the CY. This results in a
different spectrum of massless moduli on either side of the
transition, earning it the sobriquet ``drastic'' in
\cite{Greene:1996cy}.  At least in the case where the CY manifolds are
quintics in $\CM\PM^4$ the generic singularity is a conifold
singularity and as such they form an important part of the set of
singularities \cite{Greene:1995hu}.

Although such conifold points naively give a singular low energy effective
theory it is a remarkable property of string theory that they can be
understood in a well defined manner\cite{Strominger:1995cz}. When
calculating the effective action for the low energy degrees of freedom
one integrates out the heavy modes, replacing their effect by altered
couplings between the light modes. It only makes sense to integrate
out the modes which are heavier than those one is interested in. To do
otherwise is inconsistent and leads to singularities. As explained in
\cite{Strominger:1995cz} it is precisely because some unseen light
modes were being integrated out that singularities appeared in the
effective action for a conifold transition.  These extra modes
correspond to D3-branes wrapping the collapsing cycles, as the cycles
get smaller so these states become lighter and must not be integrated
out.

In this paper we examine the cosmology following from the internal CY
undergoing such a drastic topology change. The situation is rather
different to the flop where the wrapping states are massive on either
side of the flop, so that dynamically the CY will evolve to be near
the flop. For the conifold one finds that the brane wrapping states
pick up flat directions in their potential as the three-cycle turns
into a two-cycle. This creates the possibility that the CY will
dynamically evolve away from the conifold point, unlike the case of
the flop transition.  However, unless the evolution points exactly
along the flat direction the fields explore more of the potential and
are then typically forced back to the conifold point, as we shall see
in the numerical examples presented later. We also note here that a
D3-brane wrapping a three-cycle appears as a point particle in 4D, but
if it wraps a two-cycle then it gives a string like solution. Indeed,
these strings were proposed as a confinement mechanism in
\cite{Greene:1996dh}. Although it was found in \cite{Achucarro:1998er}
that such strings cannot cause confinement it does show that the flop
and the conifold transitions have rather different properties.

The paper proceeds as follows: We start by briefly describing the
properties of the conifold in section \ref{sec:conifold}. Section
\ref{sec:IIB} gives a description of how to reduce ten dimensional IIB
supergravity to four dimensions, including the effect of D3-branes.
This leads to an ${\cal N}=2$ supergravity which is presented in Section
\ref{sec:model}. The cosmological implications of this model are
presented in Section \ref{sec:cosmology} and we conclude in Section
\ref{sec:conclusions}.

\section{A brief review of the conifold geometry}
\label{sec:conifold}

\begin{figure}
\center
\epsfig{file=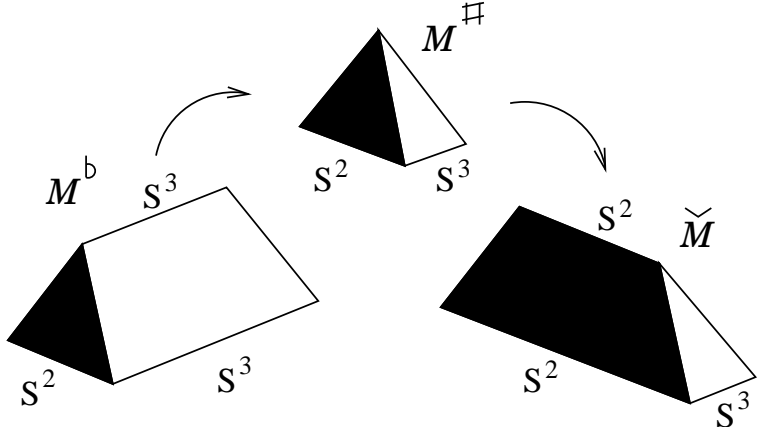,width=10cm}
\flushleft
\caption{
a representation of the conifold.
}
\label{fig:conifold}
\end{figure}

Before we discuss the low energy effective action resulting from a conifold
transition between two CYs we give here a brief introduction to the conifold
singularity, see 
\cite{Candelas:1989js}
\cite{Hubsch:1992nu}\cite{Candelas:1988di}
\cite{Candelas:1989ug}\cite{Green:1988bp}
\cite{Green:1988wa}\cite{Candelas:1990rm}
for a more complete discussion. First we review some terminology. 
For a given CY, the position of a singularity which locally looks like
the quadric in $\CM^4$
\ba
\label{eqn:coniDef}
P=\sum_{A=1}^{A=4}(W^A)^2=0
\ea
is called a {\it node}. This is the equation for a conifold singularity\cite{Candelas:1989js}. It is clearly
singular at $W^A=0$ as $P=0=dP$ at that point. We also note that it describes a 
conical shape because if $W^A$ lies on the conifold (\ref{eqn:coniDef}) then so does $\lambda W^A$.
To find the base of the cone we intersect it
with an $S^7$ of radius $r$ centred at the node,
\ba
\sum_{A=1}^{A=4}(W^A)^2=0,\qquad \sum_{A=1}^{A=4}¬|W^A|^2=r^2.
\ea
Writing $W^A=x^A+iy^A$ we find
\ba
x.x=y.y,\qquad x.y=0,\qquad x.x+y.y=r^2.
\ea
$x.x=\half r^2$ defines an $S^3$ of radius $r/\sqrt{2}$, and $y.y=\half r^2$
combined with $x.y=0$ gives an $S^2$ fibred over the $S^3$. So, the base is
$S^3$ fibred by $S^2$. As such fibrations are trivial then the base of the
conifold is the product $S^3\times S^2$. The two distinct ways for making the
conifold regular correspond to blowing up either the $S^2$ to give the
(small) {\it resolution} or by blowing up the $S^3$ to give the {\it deformation}. 
The conifold
transition then describes a CY going between these two regular manifolds.
We denote the conifold by ${\cal M}^\sharp$,
the resolved manifold by ${\cal M}^\flat$,
and the deformed manifold by $\check{\cal M}$. A nice picture of the transition
was presented in \cite{Candelas:1989js} and is given in Fig. \ref{fig:conifold}.
It shows the finite $S^3$ of the deformed conifold shrinking to zero and then
being replaced by an expanding $S^2$ of the resolved conifold. This
image will prove useful when we consider the dynamics of the transition in string
theory.

Another part of the terminology is the {\it conifold point}, this being a location
in the moduli space of CYs where the manifold acquires a node. In fact,
we shall see that for the space to remain K\"ahler it must acquire a set of
nodes \cite{Candelas:1989js}. Consider now a conifold containing $P$ such nodes
which have been deformed , thereby introducing $P$ three-cycles. Not all of these
need be homologically independent so we take there to be $Q$ homology relations
among them, giving $P-Q$ independent three-cycles. Now pick the
standard homology basis for the independent three-cycles,
\ba
A^I.B_J=\delta^I_J,\qquad I,J=0,1,2,...h^{(2,1)}.
\ea
This introduces the {\it magnetic} cycles $B_I$, dual to the {\it electric} cycles
$A^I$. 
In this work we shall consider the case where the collapsing cycles are composed
solely of electric cycles in which case, because of the $Q$ homology relations,
each $B_I$ intersects more than one collapsing cycle. Also, each vanishing cycle must
be involved in at least one homology relation if the manifold is to be K\"ahler,
as will be seen later.

\begin{figure}
\center
\epsfig{file=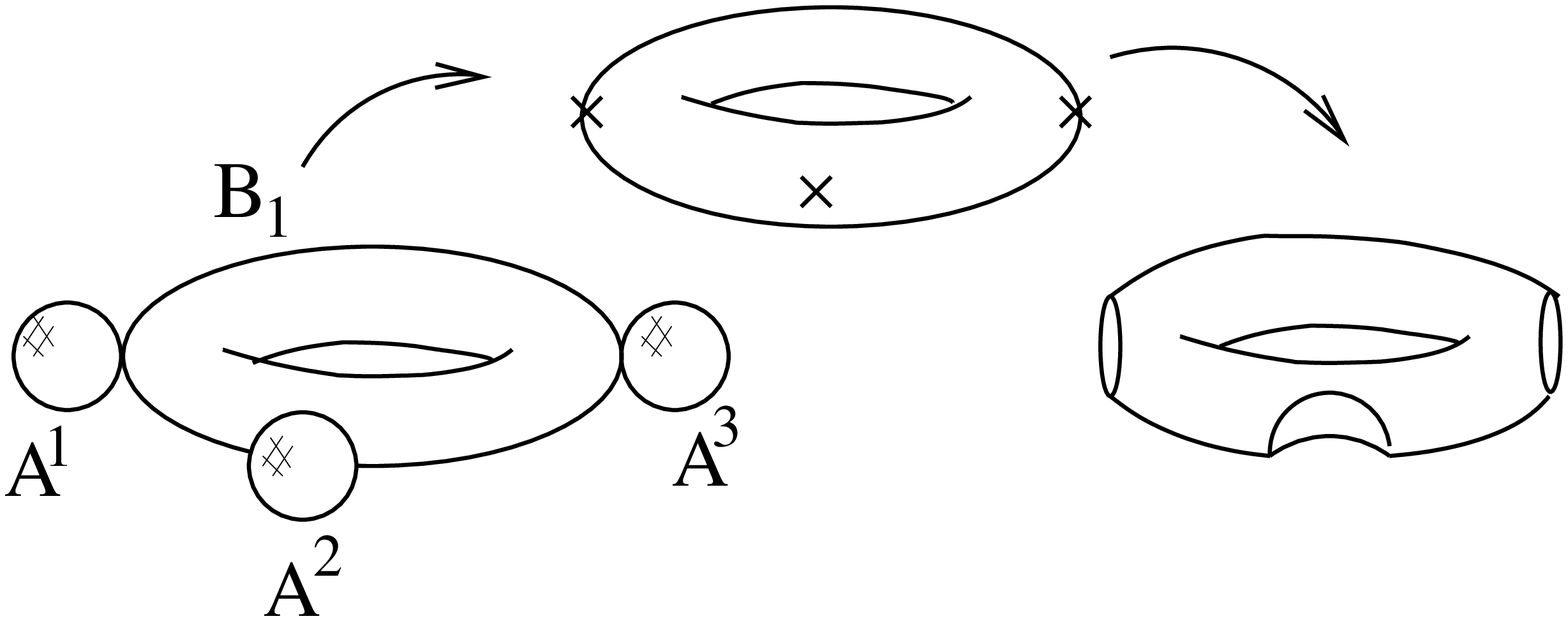,width=10cm}
\flushleft
\caption{
a representation of the cycles in a conifold transition.
}
\label{fig:cycle-chain}
\end{figure}

To see the effect of these homology relations we now 
follow a discussion of \cite{Greene:1996dh}\cite{Hubsch:1992nu} and consider the relationship
between $A$ and $B$ cycles. Fig. \ref{fig:conifold} shows a particular $A$-cycle
three-sphere, $A^1$, being blown down and then replaced by a two-sphere. The magnetic
dual of this three-cycle is constructed as follows. The shaded region is the ``cap''
$\bb{R}_{\ge 0}\times S^2$, which can be completed into a three-cycle when extended
away from the node. It is clear from the picture that this cycle intersects
$A^1$ and can be chosen as its magnetic dual, $B_1$. Also note that $B_1$ remains
a three-cycle at the node, but when the node is resolved by a two-sphere then
$B_1$ takes on the $S^2$ as a component of its boundary, so becomes a three-chain.
Each $A$-cycle that $B_1$
intersects will provide an $S^2$ component to the boundary of $B_1$ and as such
these two-spheres have a homology relation between them. Each of the magnetic
cycles provides a homology relation between the two-spheres and so we find $P-Q$
relations between the two-spheres of the resolved manifold $\check{\cal M}$.
That is, we get $P$ two-cycles with $P-Q$ homology relations, i.e. $Q$
independent two-cycles.
The picture we are left with is illustrated in Fig. \ref{fig:cycle-chain} where
the magnetic cycle $B^1$ touches the three-cycles $A^1$, $A^2$, $A^3$ which shrink
to zero and then turn into boundary two-spheres thereby converting the cycle $B^1$ into a chain.
If a $B$-cycle had intersected only one $A$-cycle then in the resolved manifold
this $B$-chain would have a single $S^2$ boundary so we would have for the
K\"ahler form $J$,
\ba
\int_{S^2}J&>&0,\\\nonumber
\int_{\del B}J&>&0,\\\nonumber
\int_{B}dJ&>&0.
\ea
Which violates the K\"ahler condition, $dJ=0$. This is why each vanishing cycle must be
involved in at least one homology relation. In particular this explains why a CY
must have more than one node\cite{Candelas:1989js}.

To re-cap then, we have that $P-Q$ independent three-spheres collapse and then expand as
$Q$ independent two-spheres. We can then see how this would appear from the point of
view of the low energy effective theory. In this picture three-cycles correspond
to complex structure moduli and each independent three-cycle gives a 4D vector-multiplet, 
while the two-cycles correspond to K\"ahler moduli and each generates
a 4D hyper-multiplet
\cite{Bodner:1989cg}\cite{Bodner:1990zm}\cite{Bohm:1999uk}\cite{Gurrieri:2003st}.
The conifold transition then sees $P-Q$ massless vector multiplets disappearing
and $Q$ massless hyper-multiplets appearing which is explained in the field theory
picture as a Higgs phenomenon \cite{Greene:1996dh}\cite{Greene:1995hu}.

\section{Reduction of IIB supergravity}
\label{sec:IIB}

In the framework of IIB supergravity we can dimensionally reduce on a CY-threefold
to find 4D ${\cal N}=2$ supergravity. The massless spectrum of this supergravity depends
on the topology of the CY with the result that there are $h^{(2,1)}$ massless
vector multiplets, $h^{(1,1)}+1$ massless hyper-multiplets and one tensor multiplet
\cite{Bodner:1989cg}\cite{Bodner:1990zm}\cite{Bohm:1999uk}\cite{Gurrieri:2003st}.
As the inclusion of D3-branes is central to the resolution of the conifold singularity
we also include the action of a probe D3-brane wrapping a three-cycle in the CY.
We start with the IIB string-frame action
given in \cite{Polchinski:1998rr}, but with the RR fields rescaled by $\sqrt{2}$
for more standard conventions in 4D. The action is given by
\ba
\label{eqn:10DIIB}
S^{10}_{IIB}&=&
   \frac{1}{2K^2_{(10)}}\int\left[e^{-2\phi}\left(R\star 1+4d\phi\wedge\star d\phi-\half H\wedge\star H\right)\right.\\\nonumber
                                &~&\left.\qquad\qquad\qquad
                                -dl\wedge\star dl-\hat F_{(3)}\wedge\star \hat F_{(3)}
                                -\half \hat F_{(5)}\wedge\star \hat F_{(5)}
                                -A_{(4)}\wedge H\wedge F_{(3)}\right],\\
\label{eqn:10DD3}
S^{10}_{D3}&=&
-\mu_3\int_{D3}d^4\xi e^{-\phi}\sqrt{-det[P(g^s_{\mu\nu})]}+\sqrt{2}\mu_3\int_{D3}A_{(4)},
\ea
where
\ba
F_{(3)}&=&dC_{(2)},\;\;\hat F_{(3)}=F_{(3)}-lH,\;\;\hat F_{(5)}=dA_{(4)}-H\wedge C_{(2)},\;\;
\mu_3=\sqrt{\pi}/K_{(10)}.
\ea
The equation of motion and Bianchi identity for $\hat F_{(5)}$ are
\ba
d\star\hat F_{(5)}&=&H\wedge F_{(3)}=d\hat F_{(5)},
\ea
so that we may consistently impose the required self-duality of the five-form field strength
$\hat F_{(5)}=\star \hat F_{(5)}$.

We shall only be interested in the case where $H$, $l$ and $C_{(2)}$ vanish. Then we can follow
the standard procedure of expanding $A_{(4)}$ in terms of harmonic forms living on the CY,
\ba
A_{(4)}&=&V^I\wedge\alpha_I-U_I\wedge\beta^I,\\
\hat F_{(5)}&=&F^I\wedge\alpha_I-G_I\wedge\beta^I,
\ea
where we have introduced a basis of harmonic three-forms $\alpha_I$, $\beta^I$ 
and the one-forms $V^I$, $U_I$ live on the spacetime and correspond to Abelian
gauge fields with field strengths $F^I=dV^I$, $G_I=dU_I$.
The harmonic forms are normalized to satisfy
\ba
\int_{A^I}\alpha_J=\int_{B_J}\beta^I=\sqrt v\delta^I_J,\;\;\;\;
\int_{CY}\alpha_I\wedge\beta^J=v\delta_I^J
\ea
with $v$ being some reference CY volume.
Self duality of $\hat F_{(5)}$ then relates $G_I$ to $F^I$ by
\ba
\label{eqn:magField}
G_I&=&(ReM)_{IJ}F^J+(ImM)_{IJ}\star F^J,
\ea
where we 
define the matrices $A$, $B$, $C$, $D$, $M$ \cite{Gurrieri:2003st}
\ba
\star\alpha_I&=&A_I^{\;\;J}\alpha_J+B_{IJ}\beta^J,\\
\star\beta^I&=&C^{IJ}\alpha_J+D^I_{\;\;J}\beta^J,\\
\label{eqn:M}
A&=&(ReM)(ImM)^{-1},\\\nonumber
B&=&-(ImM)-(ReM)(ImM)^{-1}(ReM),\\\nonumber
C&=&(ImM)^{-1}.
\ea
The relation between $G_I$ and $F^I$ has important consequences for the interpretation
of the charges of the particle states coming from wrapped branes. We shall see
shortly that $G_I$ is in fact the magnetic dual of $F^I$, so states charged under
$G_I$ are in fact magnetic charges of $F^I$.

With this decomposition of the field strengths we are in a position to dimensionally
reduce the D3-brane action (\ref{eqn:10DD3}).
To do so one first changes to the 10D Einstein frame and writes
the metric in this frame as a direct product between a CY and a 4D spacetime. This gives
an action in 4D which needs to be Weyl rescaled by the volume of the CY,
$\kappa$, to get it in the 4D Einstein frame,
\ba
\kappa=\frac{1}{6}\int_{CY}J\wedge J\wedge J.
\ea
One then finds that\cite{Gurrieri:2003st}
\ba
\label{4DIIB}
S^{4}_{IIB}&=&
       \frac{1}{K^2_{(4)}}\int\left[ \half R\star 1
                                   -g_{i\bar j}dz^i\wedge\star dz^{\bar j}
                                   -h_{uv}dq^u\wedge\star dq^{v}\right.\\\nonumber
                                   &~&\qquad\qquad\left.+\half F^I\wedge\left(Im(M)_{IJ} \star F^J
                                   +Re(M)_{IJ}F^J\right)\right],\\
K_{(4)}^2&=&K^2_{(10)}/v,\\
i,j&=&1,2...h^{2,1},\;\;I,J=0,1...h^{2,1},
\ea
where the complex structure moduli, $z^i$, have a moduli metric $g_{i\bar j}$ derived from
the K\"ahler potential $K_{(\Omega)}$,
\ba
e^{-K_{(\Omega)}}&=&(i/v)\int_{CY}\Omega\wedge\bar\Omega=||\Omega||^2\kappa/v,
\ea
and the fields $q^u$ are the hyper-multiplets which have a
quaternion-K\"ahler moduli metric $h_{uv}$. 
Written in this form, (\ref{4DIIB}) shows that the field strength
$G_I$ (\ref{eqn:magField}) is the magnetic dual of $F^I$ as was claimed earlier.

Along with the reduction of the supergravity we also need to dimensionally reduce
the D3-brane action (\ref{eqn:10DD3}), but to do that we need to know more about the cycle it is
wrapping. Quite generally we may decompose this cycle ${\cal C}$ as
\ba
{\cal C}&=&n_I A^I+m^I B_I,\qquad n_I,\;m_I\;\in \bb{Z}.
\ea
The extra piece of information which we need is that the D3-brane 
lives on a minimal, supersymmetric
SLAG three-cycle \cite{Joyce:2001nm}\cite{Becker:1995kb}\cite{Vafa:1995ta}.
Such cycles are calibrated by $Re(e^{i\theta}\Omega)$, for some constant $\theta$,
and saturate the following bound
\ba
Vol({\cal C})\geq\frac{\sqrt\kappa\left|\int\Omega\right|}{\left|\int\Omega\wedge\bar\Omega\right|}
      =\sqrt{(\kappa/v)}e^{(K_{(\Omega)}/2)}\left|\int\Omega\right|.
\ea
By writing the holomorphic three-form as 
\ba
\Omega&=&X^I\wedge\alpha_I-{\cal F}_I\wedge\beta^I,
\ea
we find 
\ba
e^{-K_\Omega}&=&\frac{i}{v}\int\Omega\wedge\bar\Omega=i\left({\cal F}_I\bar X^I-X^I\bar {\cal F}_I\right),
\ea
and by wrapping the brane on a SLAG cycle we can perform the spatial integration in (\ref{eqn:10DD3})
to find that the D3-brane action becomes, after the Weyl re-scalings,
\ba
\label{4DD3}
S^{4}_{D3}&=&-\frac{\sqrt{\pi}}{K_{(4)}}e^{(K_\Omega/2)}|n_IX^I-m^IF_I|\int d\tau
                                   +\frac{\sqrt{2\pi}}{K_{(4)}}n_I\int V^I
                                   -\frac{\sqrt{2\pi}}{K_{(4)}}m^I\int U_I.
\ea
This corresponds to a particle of mass $\frac{\sqrt{\pi}}{K_{(4)}}e^{(K_\Omega/2)}|n_IX^I-m^IF_I|$,
charged under the gauge fields $U_I$ and $V_I$. Here, $\tau$ is the proper time of the particle.
This relation between mass and charge is what is to be expected for an ${\cal N}=2$ supergravity
\cite{Ceresole:1995ca}.
As explained earlier, electric charges of $U_I$ are in fact magnetic charges of $V^I$. This
explains why the $A^I$ and $B_I$ cycles were termed electric and magnetic respectively.
We shall only be interested in cases where the particles have electric charge
so we set $m^I$ to zero and only wrap the D3-brane over the $A^I$.

\section{${\cal N}=2$ supergravity}
\label{sec:model}
The way we proceed is to use the structure of ${\cal N}=2$ supergravity along
with the information we have about the extra brane wrapping states to write
down the 4D effective action of the conifold transition. We still have $h^{(2,1)}$
vector multiplets but now there are $h^{(1,1)}+1+P$ hyper multiplets, corresponding
to the usual set plus $P$ from wrapped three-cycles. Due to the $Q$ homology
relations these extra hyper multiplets are charged under $P-Q$ of the vector
fields. What we require then is a gauged supergravity to account for
hyper-multiplets charged under Abelian gauge groups.
Taking results from \cite{Andrianopoli:1996cm}\cite{Andrianopoli:1996vr}
and using our conventions we have 
\ba
\label{N=2action}
S^{4}_{{\cal N}=2}&=&
       \frac{1}{K^2_{(4)}}\int\left[ \half R\star 1
                                   -g_{i\bar j}dz^i\wedge\star dz^{\bar j}
                                   -h_{uv}Dq^u\wedge\star Dq^{v}\right.\\\nonumber
                                   &~&\qquad\qquad+\half Im({\cal N})_{IJ}F^I\wedge \star F^J
                                   -\half Re({\cal N})_{IJ}F^I\wedge F^J\\\nonumber
                                   &~&\qquad\qquad\left.-g^2\left[ 4h_{uv}k^u_I k^v_J\bar L^IL^J
                                             +(g^{i\bar j}f^I_i f^J_{\bar j}
                                             -3\bar L^I L^J){\cal P}^x_I{\cal P}^x_J\right]\right],\\
Dq^u&=&dq^u+gA^Ik^u_I(q),\\
L^I&=&e^{(K_{(\Omega)}/2)}X^I,\\
{\cal F}_I&=&{\cal N}_{IJ}X^J,\\
f_i^I&=&\left(\partial_i + \half \partial_i K_{(\Omega)} \right)e^{\half K_{(\Omega)}} X^I,\\
f_{\bar{i}}^I&=&\left(\partial_{\bar{i}} + \half \partial_{\bar{i}} K_{(\Omega)} \right)e^{\half K_{(\Omega)}} \bar{X}^I\\
\label{KahlerPotential}
\exp(-K_{(\Omega)})&=&i(\bar{X}^I{\cal F}_I-X^I\bar{\cal F}_I),\\
\label{KillingPrePot}
k_I\lrcorner K^x&=&\nabla{\cal P}^x_I.
\ea
The $X^I$ are holomorphic functions of the $z^i$ and 
if we use special co-ordinates then we have that the complex structure part of the
action is determined by a pre-potential, ${\cal F}$, which is a degree two
holomorphic function of $X^I$. The K\"ahler potential (\ref{KillingPrePot})
is then determined in terms of the first derivatives
\ba
{\cal F}_I&=&\frac{\del {\cal F}}{\del X^I},
\ea
and can be obtained as a function of $z^i$ by setting $X^I=(1,z^i)$.
The matrix ${\cal N}_{IJ}$ is known as the period matrix and is given by 
\cite{Ceresole:1995ca}\cite{Andrianopoli:1996cm}\cite{Andrianopoli:1996vr}
\be
{\cal N}_{IJ} = \bar{{\cal F}}_{IJ} + \frac{2i(Im {\cal F}_{IK})X^K (Im {\cal F}_{JL})X^L}{(Im {\cal F}_{MN})X^MX^N}
\ee
where
\be
{\cal F}_{IJ} = \partial_I \partial_J {\cal F}
\ee
Comparison of action
(\ref{N=2action}) with (\ref{4DIIB}) shows that this period matrix equates to the matrix
$M_{IJ}$ introduced in (\ref{eqn:M}).
This action contains two important metrics, $g_{i\bar j}$ and $h_{uv}$. The gauging of
the scalars comes by gauging with respect to isometries of these metrics.
Here we are only interested in giving the hyper-multiplets charge with respect
to an Abelian group and so we have restricted the action of 
\cite{Andrianopoli:1996cm}\cite{Andrianopoli:1996vr}
to give the above form, with $k^u_I$ being the Killing vectors
on $h_{uv}$. Further, ${\cal P}^x_I$ are the associated pre-potentials
and the triplet of two-forms $K^x$ 
represent the hyper-K\"ahler forms. The operator $\nabla$ appearing in (\ref{KillingPrePot}) is the SU(2)
covariant derivative on $h_{uv}$. 

\subsection{An example}
\label{sec:example}

To make any progress we need to pick a particular CY to reduce on. This is equivalent
to choosing a pre-potential for the complex structure and a quaternionic metric for
the hyper-multiplets.
The exact quaternionic sigma model manifold for the brane wrapping hypermultiplets 
is unknown as yet and will in general mix the wrapping-hypermultiplet
components with the K\"ahler moduli. This is necessary because the overall hypermultiplet
manifold must be quaternionic, and as such cannot be a direct product of 
manifolds \cite{Jarv:2003qx}.
However, we know from the brane reduction that the mass term for the hypermultiplets only involves complex 
structure terms and not the K\"ahler moduli, this 
means that to leading order we can neglect the coupling between the K\"ahler moduli and
the hypermultiplets representing the wrapping states.
As such, as long as the hypermultiplets remain small we can consider the approximation
that the quaternionic metric is flat and switch off all hypermultiplets other than the wrapping states.
Higher order terms will couple the brane 
wrapping states to the usual hyper-multiplets forming the full quaternionic manifold.
However, we do not expect the higher order terms to alter the qualitative behaviour of the dynamics.     
For example,the initial study of \cite{Brandle:2002fa} on flop transitions made this same approximation
which was amended in \cite{Jarv:2003qy}\cite{Jarv:2003qx} but lead to the same structure. More recently, dynamics of 
conifolds transitions in the context of M-theory were examined with both a first order approximation and using 
complete Wolf spaces \cite{Mohaupt:2004pr}; the same conclusion was reached.  
With this understood we will set $h_{uv}=\delta_{uv}$ and so the $\nabla$ in (\ref{KillingPrePot}) gets replaced 
by the usual exterior derivative. The simulations in the following section will reveal that 
the $q^{au}$ remain small in our examples, maintaining the validaty of our approximation.

Now we must construct the Killing vectors
which are used to gauge the hyper-multiplets.
We shall denote the hyper-multiplets by indices $a,b=1,2,...P$ where 
we have switched off the other $h^{(1,1)}+1$ hyper-multiplets which come from
the usual supergravity reduction.
This is consistent with the first order approximation we are using for the hyper-multiplet metric.
The components of a given hyper-multiplet
are labelled with the indices $u,v=1,2,3,4$ so we write $q^{au}$ for hyper-multiplet components,
their charge is given in units of $g$ as $gQ^a_I$.
The gauge fields take on the labels $I,J=0,1,...h^{(2,1)}$ but for $Q$ homology
relations among the $P$ wrapped cycles only $P-Q$ of these gauge fields are used
in the gauging of $q^{au}$.
As our metric is taken to be flat we have a choice of rotation or translation
Killing vectors. For the potential to vanish when the $q^{au}$ do then the rotation
Killing vectors are the relevant choice,
\ba
k^{au}_I=\sum Q^a_I t^u_{\;\;v}q^{av},
\ea
where $t$ is an anti-symmetric matrix. Moreover, we can introduce the 't Hooft symbols
\cite{'tHooft:1976fv}.
\ba
\eta^a_{bc}=\bar\eta^a_{bc}=\epsilon^a_{\;bc},\\
\eta^a_{b0}=\bar\eta^a_{0b}=\delta^a_b,
\ea
and write a set of hyper-K\"ahler forms as
\ba
K^x_{au,bv}=-\bar\eta^x_{uv}\delta_{ab}.
\ea
By decomposing the matrix $t$ as 
\ba
t_{uv}&=&n_x\eta^x_{uv},
\ea
where $\underline n$ is a unit vector, we find the following pre-potential
for the Killing vectors,
\ba
\label{eqn:KillingPrePot}
{\cal P}^x_I&=&\half\sum Q^a_Iq^{av}\left(\bar\eta^x_{uv}n_y\eta^{y,u}_{\;\;w}\right)q^{aw}.
\ea
Here we have dropped the possible constants of integration (Fayet-Illiopoulos terms)
as they would give a potential for the complex structure even in the absence of brane
wrapping i.e. $q^u=0$.
Then the Killing vector terms in the potential become
\ba
\label{eqn:Dterm}
V^{(D)}_{IJ}&=&P^{x}_{I}P^{x}_{J} = \sum Q^{a}_{I}Q^{b}_{J} 
                           \left( \half q^{av}q^{aw}q^{bv}q^{bw} 
                                 -\quarter q^{av}q^{av}q^{bw}q^{bw} 
                                 +\half q^{av}q^{aw}q^{br}q^{bt}t_{wt}t_{vr} \right),\\
\label{eqn:mterm}
V^{(m)}_{IJ}&=&h_{au,av}k^{au}_{I}k^{av}_{J} = \sum Q^{a}_{I}Q^{a}_{J}q^{aw}q_{aw}.
\ea
The quadratic part of the hyper-multiplet Lagrangian may then be extracted
to find the their mass in terms of $g$, then using the earlier result that branes wrapped
on electric cycles have mass $\frac{\sqrt{\pi}}{K_{(4)}}e^{(K_\Omega/2)}|n_IX^I|$
we conclude that
\ba
g^2&=&\pi/(4K^2_{(4)}).
\ea

To fix the rest of the model we need to choose a pre-potential ${\cal F}$, to do
this we take the generic large complex structure form
\ba
{\cal F}&=&d_{IJK}\frac{X^IX^JX^K}{X^0} \label{Flarge},
\ea
for constant $d_{IJK}$,
and restrict to the more manageable case
\ba
\label{eqn:periodChoice}
d_{IJ0}=-\frac{i}{2}N_{IJ}, \qquad N_{0i}=0, \qquad d_{IJi}=0 
\ea
where we have introduced the constant coupling matrix $N_{IJ}$. 
Of course, the above form of the pre-potential is generically invalidated
as some of the three-cycles become small near the conifold transition. 
However, corrections to the pre-potential will come in the form of logrithmic terms originating from the monodromy about the conifold
point and other terms which are analytic at the conifold point\cite{Strominger:1995cz}. 
The logarithmic terms are accounted for through the inclusion
of the states that become light, and corrections which are not of the form (\ref{eqn:periodChoice}) will be cubic 
and higher order; these will be negelible very close to the conifold point where the periods are vanishing. Hence we expect
that they will not affect our qualitative conclusions and as a first 
approximation we do not include them.  
With this understood, we find that our 4D effective action becomes
\ba
\label{completehyperaction}
S_{IIB+D3}^{(4)} = \nonumber \frac{1}{K^2_{4}} \int \sqrt{-g}d^4x \left\{\;\; \half R_4\right.
                &-& \left(\frac{-N_{ij}}{<X|X>} + \frac{z^{m}\bar{z}^n N_{mj} N_{ni}}{{<X|X>}^2} \right)
                   \partial_{\mu}z^{i}\partial^{\mu}\bar{z}^j \\ \nonumber
                &-& \nabla_{\mu}q^{au}\nabla^{\mu}q_{au} 
                - 2g^2 \left( \frac{z^i\bar{z}^j}{<X|X>} \right) 
                           V^{(m)}_{ij}  \\ \nonumber
                &-& g^2 \left( - \half (N^{-1})^{ij} - \frac{\bar{z}^i z^j}{<X|X>}
                          \right) V^{(D)}_{ij}  \\  
                &-& \left.\frac{1}{4} Im\left({\cal N}\right)_{IJ}F^I_{\mu\nu} F^{J\mu\nu}\right\}.
\ea
where
\be
<X|X> =  N_{IJ} X^I \bar{X}^J = N_{00} + N_{ij}z^i\bar{z}^j
\ee
One finds that in order to have positive kinetic terms for the scalar fields while still satisfying (\ref{KahlerPotential}) the coupling matrix 
$N_{IJ}$ must have signature (+,-,-,...)\cite{Ceresole:1995ca}\cite{Andrianopoli:1996cm}\cite{Andrianopoli:1996vr}.

\subsection{Cosmology of conifolds.}
\label{sec:cosmology}

As we are interested in the cosmological evolution following from a conifold transition
we shall take all the scalar fields to to depend only on time. The vector fields vanish
by isotropy considerations. If we take the spacetime metric to Friedmann-Robertson-Walker
with flat spatial sections,
\ba
ds^2=-dt^2+a(t)^2 d\underline{x}^2,
\ea
then we find the following equations of motion from (\ref{completehyperaction}).
\ba
\ddot{q}_{av} + 3\left(\frac{\dot{a}}{a}\right)\dot{q}_{av}  + \half\partial_{a}V &=& 0,\\
\label{eqn:zeqn}
\ddot{z}^i + 3\left(\frac{\dot{a}}{a}\right)\dot{z}^i + \Gamma^i_{\;\;jk}\dot{z}^j\dot{z}^k 
+ \partial_{\bar j} Vg^{\bar j i} &=&0,\\
\label{eqn:chargeConstraint}
Q_I^a t^u_{\;\;v} q^{av} \partial_{\mu}q_{au}&=&0,\\
2\left(\frac{\ddot{a}}{a}\right) + \left(\frac{\dot{a}}{a}\right)^2 &=& -g_{i\bar j}\dot{z}^i\dot{\bar{z}}^j 
-\dot{q}^{av}\dot{q}_{av} + V,\\
\label{eqn:FRWConstraint}
3\left(\frac{\dot{a}}{a}\right)^2 &=& g_{i\bar j}\dot{z}^i\dot{\bar{z}}^j + \dot{q}^{av}\dot{q}_{av} + V
\ea
where 
\be
V = 2g^2 \left( \frac{z^i\bar{z}^j}{<X|X>} \right)  V^{(m)}_{ij}
    + g^2 \left( - \half (N^{-1})^{ij} - \frac{\bar{z}^i z^j}{<X|X>} \right) V^{(D)}_{ij}
\ee
The connection on the complex structure moduli space is given by
\be
\Gamma^i_{\;\;jk}=g^{\bar l i}\partial_j g_{k\bar l}.
\ee
When performing the numerical simulation, the gauge field equation (\ref{eqn:chargeConstraint}) and the energy
conservation equation (\ref{eqn:FRWConstraint}) act as a check that the simulations are accurate. 
Also note that (\ref{eqn:chargeConstraint}) is expressing the statement that there are no electric
currents. In particular, for the cosmological simulations we have zero charge density, which corresponds
to no net brane wrapping in the string theory picture.
We shall perform our simulations in units where
$\sqrt\pi / (2K_{(4)})=1$.

\begin{figure}
\center
\epsfig{file=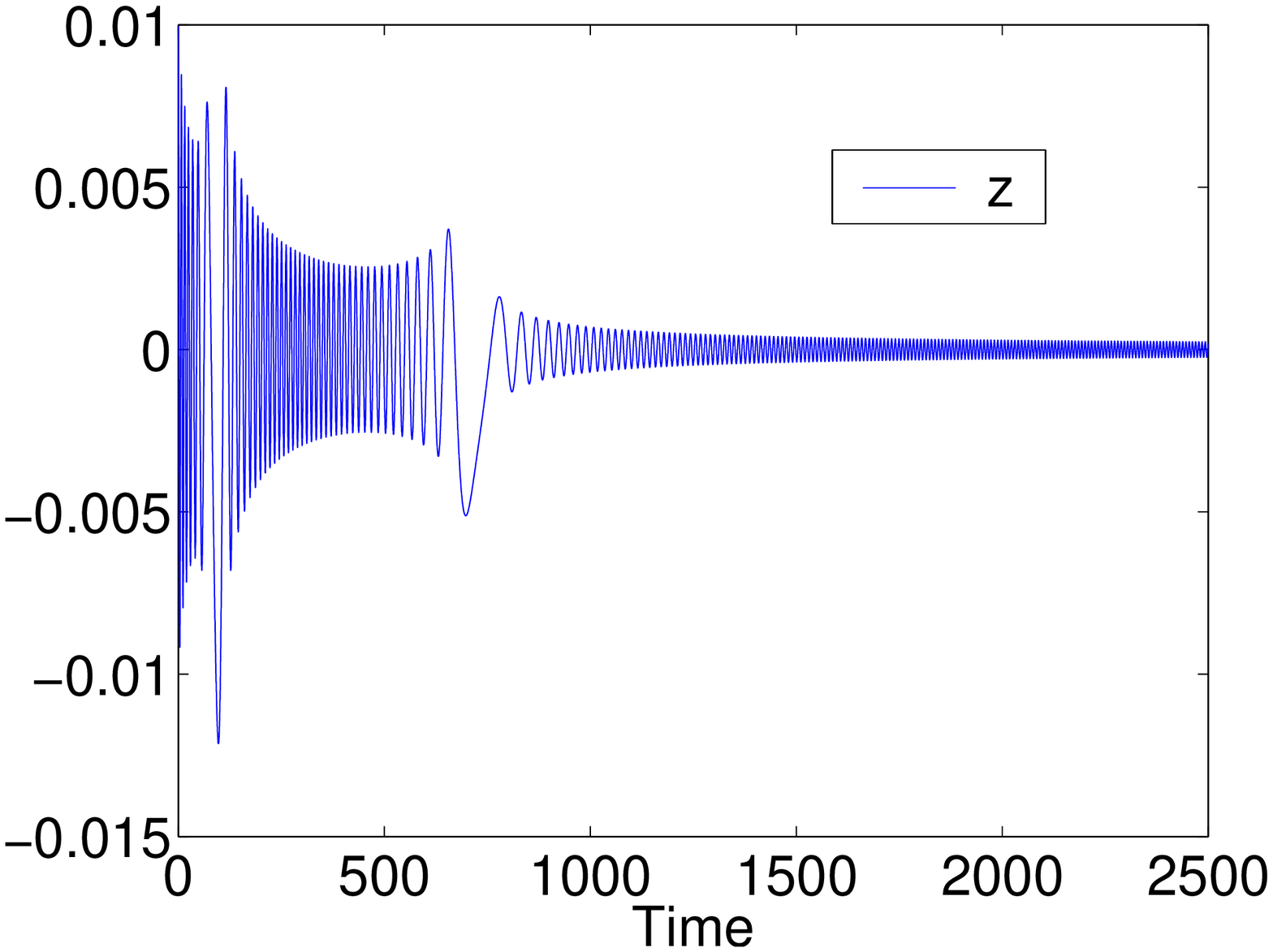,width=7.5cm}
\epsfig{file=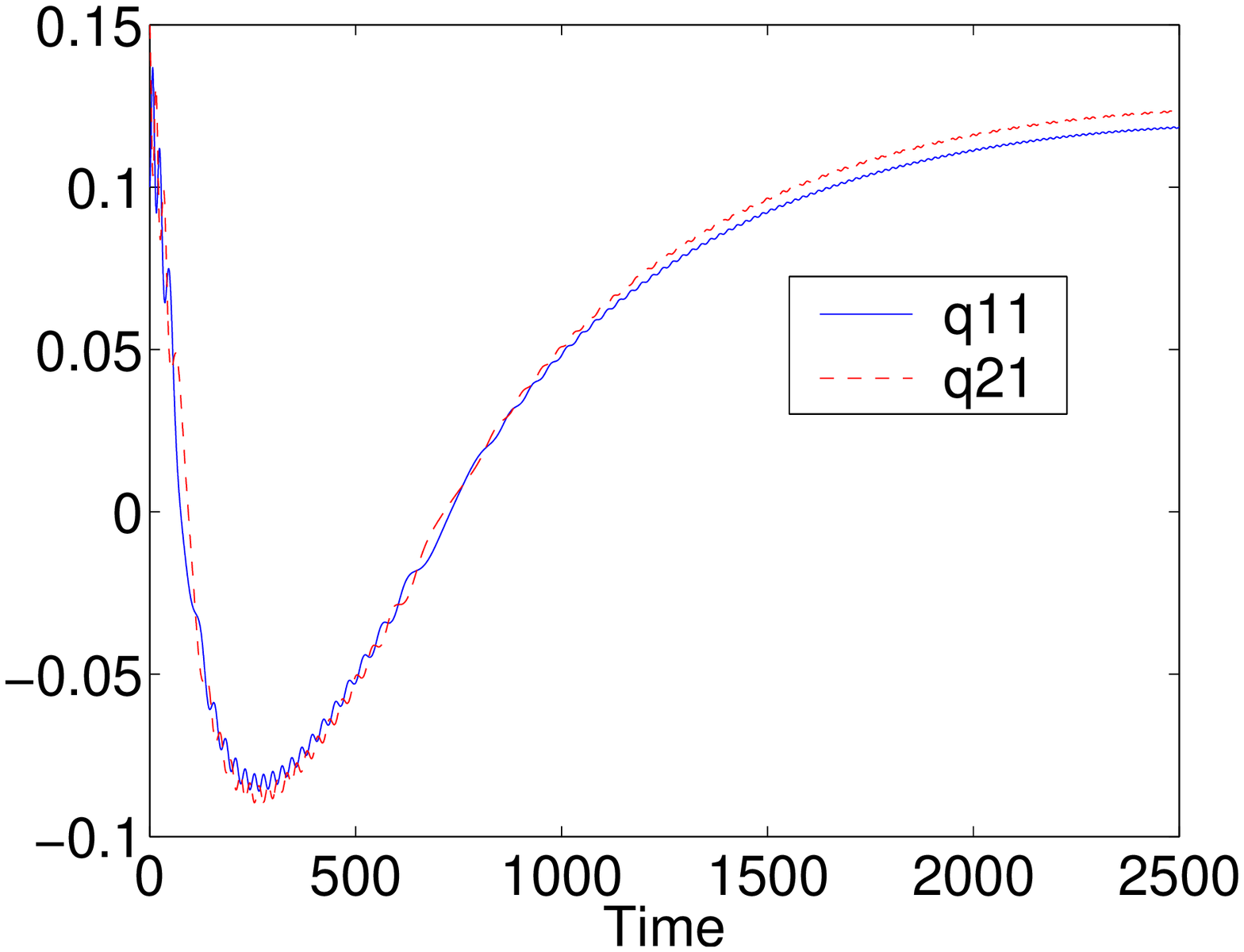,width=7.5cm} \\
\epsfig{file=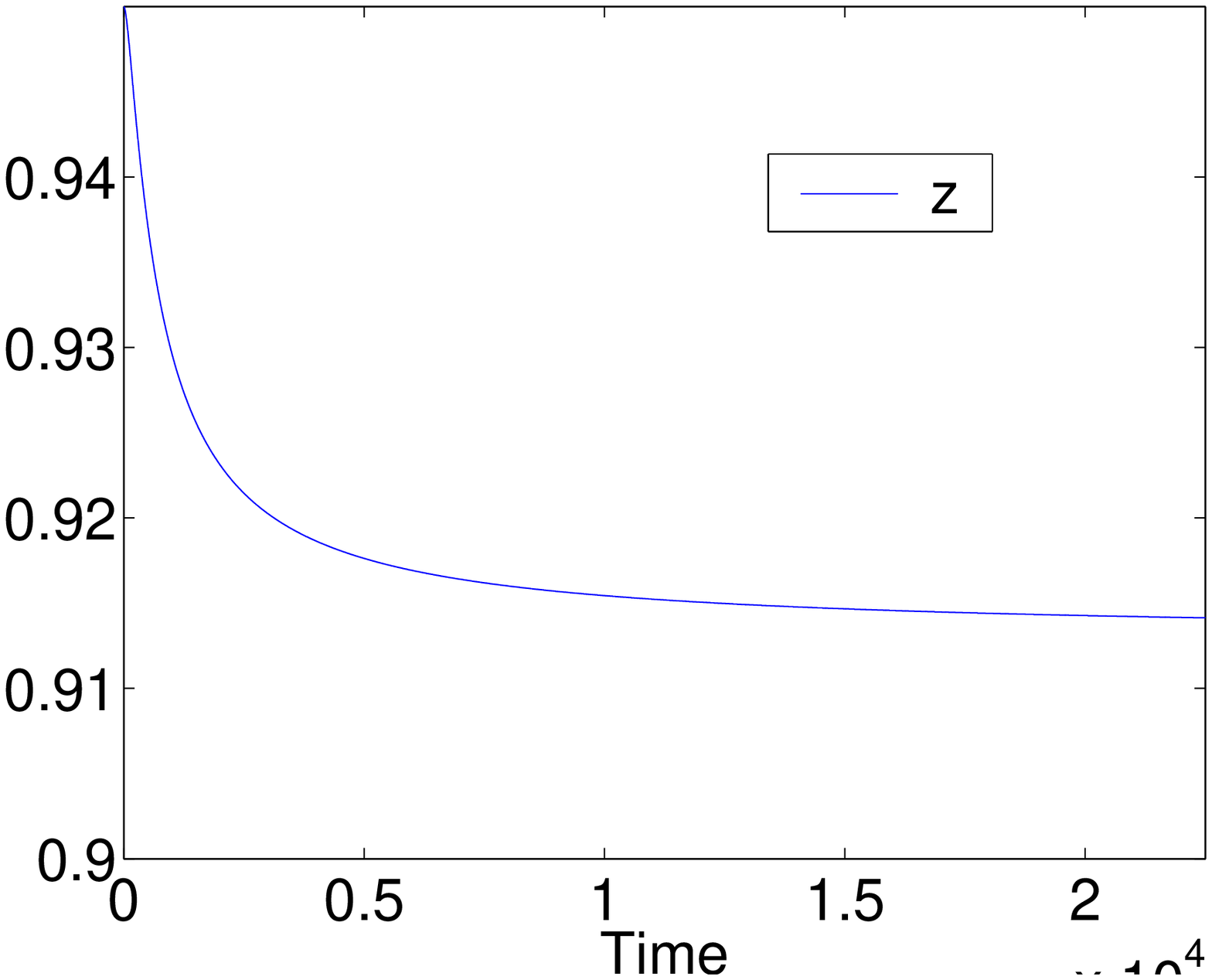,width=7.5cm}
\epsfig{file=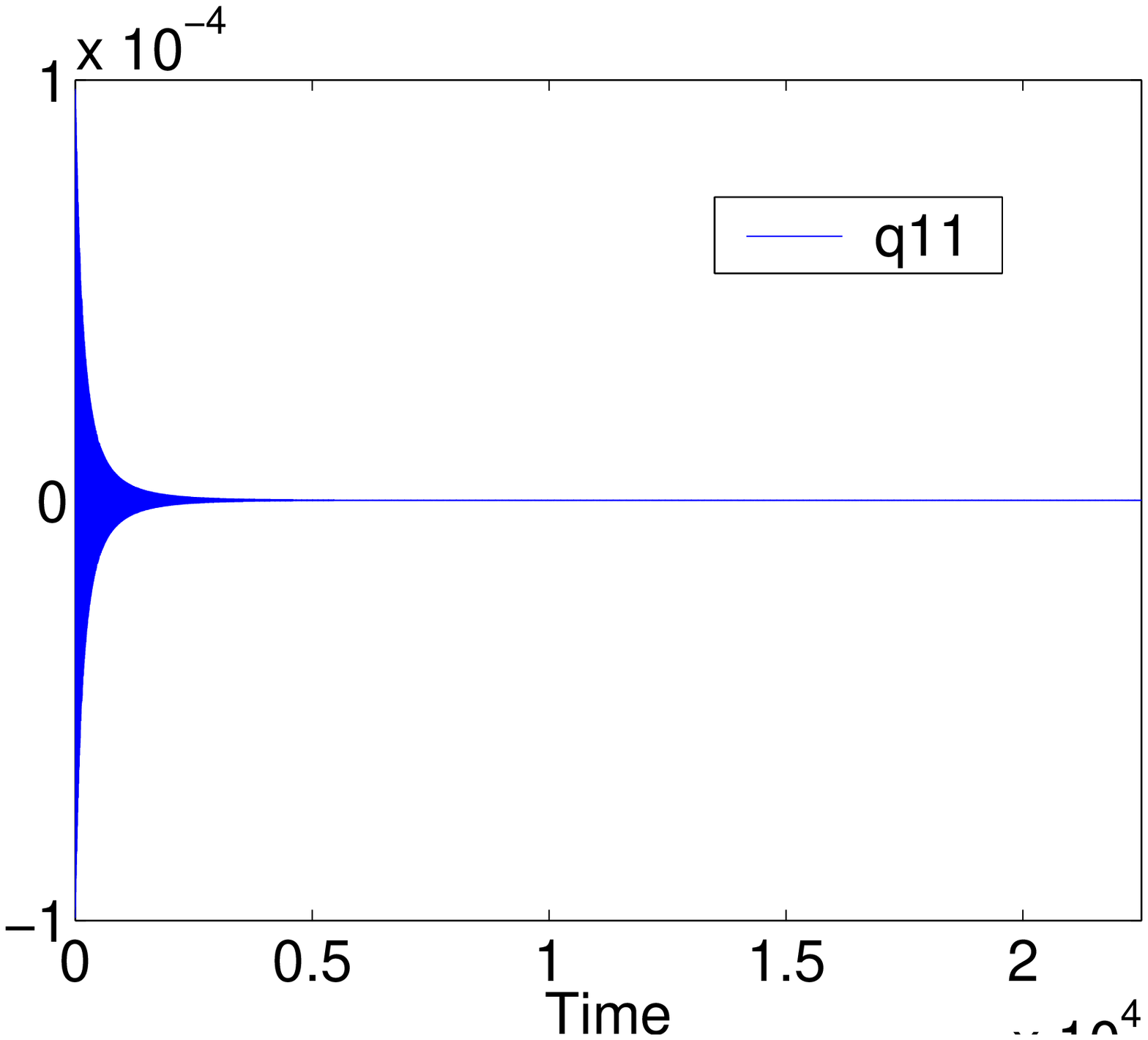,width=7.5cm}\\
\epsfig{file=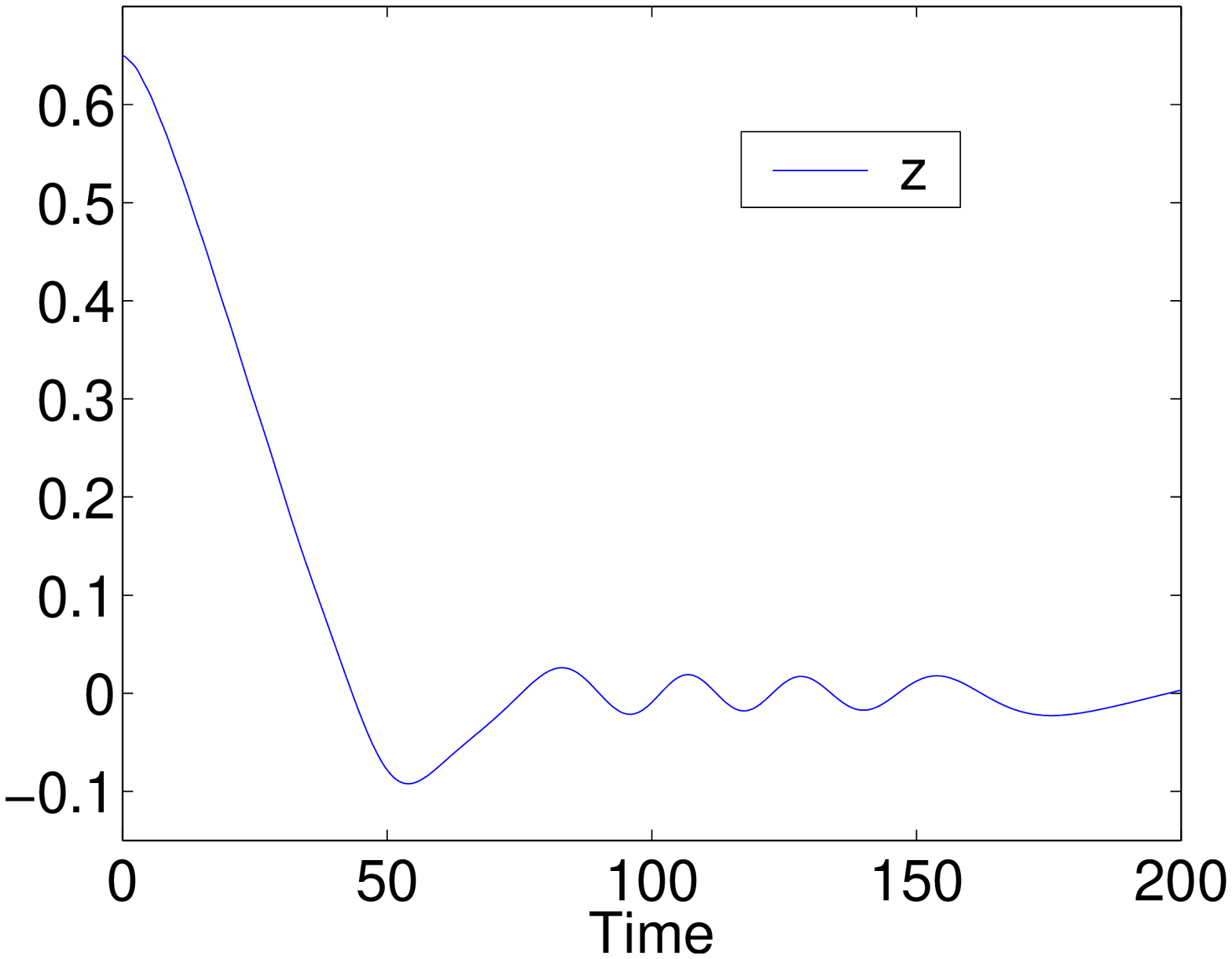,width=7.5cm}
\epsfig{file=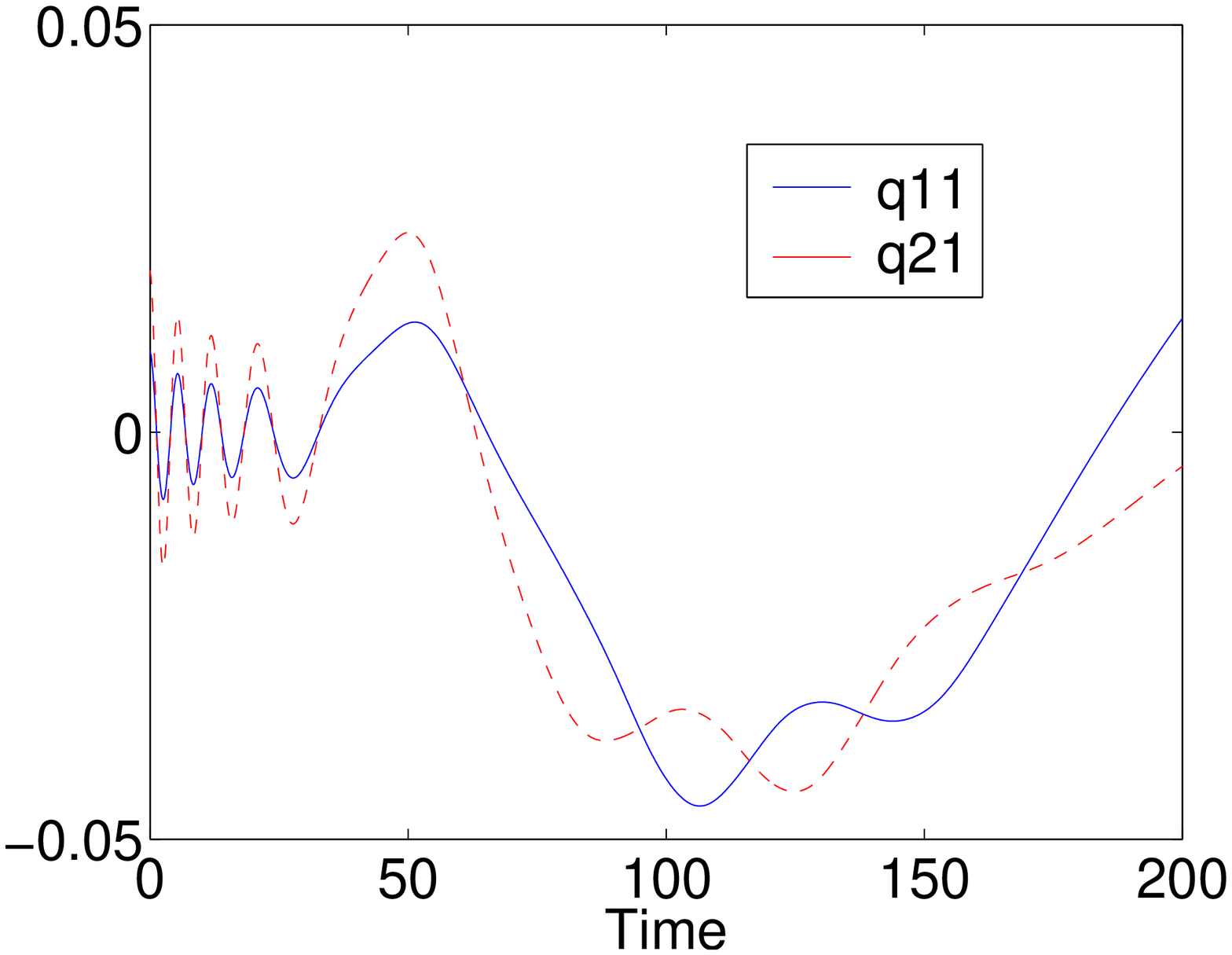,width=7.5cm}
\flushleft
\caption{Figure showing the evolution of the fields $Re(z)$, $q^{11}$ and $q^{21}$ against time.}
\label{fall}
\end{figure}

An interesting effect with the flop transition,
noted in \cite{Brandle:2002fa}, is that even the moduli which are not involved
in the flop were found to be stabilized once the additional hyper-multiplets
acquire a non-zero value. 
Here we note a similar effect
is possible depending on the choice for the coupling matrix $N_{IJ}$. To see this let us
consider the equation of motion for the complex structure (\ref{eqn:zeqn}) using (\ref{eqn:periodChoice})
\ba
\label{eqn:unwrapped}
\ddot{z}^i + 3\left(\frac{\dot{a}}{a}\right)\dot{z}^i + \Gamma^i_{\;\;jk}\dot{z}^j\dot{z}^k
- \left( N^{-1} \right)^{ij}z^k\left( V^{(m)}_{jk} -  V^{(D)}_{jk}  \right) = 0.
\ea
As hyper-multiplets are only charged under the gauge fields associated with the wrapped three-cycles,
then one can easily show that those $z^i$ where the index $i$ corresponds to an unwrapped cycle
can be a flat directions in the potential; $\del_{\bar j}g^{\bar j i} V=0$. 
To see this, note from (\ref{eqn:Dterm})(\ref{eqn:mterm}) that $V^{(m)}_{jk}$, $V^{(D)}_{jk}$
vanish for those indices $j,k$ not associated to the charges. 
A particularly simple choice for $N_{IJ}$ is minimal coupling, with $N_{IJ}=\eta_{IJ}$
\cite{deWit:1984pk} giving
\be
N_{00} = 1 \;\;\;\; N_{i0} = 0 \;\;\;\; N_{ij} = -\delta_{ij}.
\ee
For this case (\ref{eqn:unwrapped})
gives a vanishing $\del_{\bar j}g^{\bar j i} V$ for those $i$ which have $Q^a_i=0$.
We could say that in the case of minimal
coupling only the wrapped $z^i$ have a potential. If we think of the period matrix as the means
by which the different vector multiplets couple to each other then allowing off diagonal terms
in $N_{ij}$ lets the unwrapped $z^i$ communicate with the wrapped ones, so in this case if the
moduli for the wrapped cycles get stabilized there is also the possibility that the moduli for
unwrapped cycles will get stabilized. 
Indeed, from (\ref{eqn:unwrapped}) we see that for non-minimal coupling (allowing off-diagonal
terms in $N_{ij}$) even the $z^i$ which are unwrapped ($Q^a_i=0$) have a forcing term which
is linear in $z$, corresponding to a quadratic potential.
In our numerical experiments we present an example for a particular non-minimal
period matrix where all of the $z^i$ get stabilized.
It is important to note that at the conifold point, or when no cycles are wrapped, the potential for all the moduli vanishes. 
Similarly on the Higgs branch only the wrapped moduli have a potential. And so we see that eventually the unwrapped moduli will
return to be flat directions, but with an increased probability of being found near the location where the 
potential had a minimum during the transition.
If both 
$z^i$ and $q^{au}$ are non-zero then they generate an effective mass for each other and so each are driven
to oscillate about zero.
Because of the cosmological expansion these oscillations are damped and we have several possibilities:
The $z^i$ could reach zero first leaving the $q^{au}$ to wander in their flat directions
and forcing the field theory into the Higgs phase; the $q^{au}$ could vanish first in which case the
$z^i$ will not be driven to zero and the field theory remains in the Coulomb phase; or both the complex
structure and the hyper-multiplets go to zero and the CY remains near the conifold point. Which of these
possibilities actually occurs clearly depends on the initial conditions of the fields and we present
numerical examples of all three scenarios.
Note that it is consistent for the $z^i$ to remain non-zero while the $q^{au}$ decay because we
have zero net brane wrapping, and so, we believe, the decay of $q^{au}$ while the three-cycle is non-vanishing can
be interpreted as the annihilation of brane anti-brane pairs. 
The initial values for the fields
are shown in Table \ref{tall} and we take vanishing time derivatives for all the scalars as initial conditions.
The evolution of the fields $z$, $q^{11}$, and $q^{21}$ is plotted in Fig. \ref{fall}. The scale factor
has not been plotted as this just increased monotonically
from its initial value of unity. Neither did we plot all of the components of the hyper-multiplets,
these behaved in a similar manner to the first component so, to keep the plots simple, they are not shown.

For our first simulation we shall consider the case where the CY has two collapsing three-cycles
such that their sum is homologically trivial. In the field theory this corresponds to two hyper-multiplets,
$q^{1u}$ and $q^{2u}$,
charged under a single gauge field $A^1$ with opposite charges, which we take as $Q^1_1=1$, $Q^2_1=-1$. We can now see explicitly
how the flat directions appear once the three-cycles have collapsed, $z^1 \equiv z=0$, by examining the potential
in (\ref{N=2action}). The only term which survives is the term quadratic in the pre-potentials ${\cal P}^x_I$,
and in the case of minimal coupling this term is $~\delta^{ij}{\cal P}^x_i{\cal P}^x_j$ \footnote{${\cal P}^x_0 = 0$}
which is positive definite and vanishes when ${\cal P}^x_i$ does. By examining (\ref{eqn:KillingPrePot}) we see then that
for $q^{1u}=q^{2u}$ then ${\cal P}^x_i=0$, so our flat direction drives the components of the two hyper-multiplets
towards each other. It is also trivial to see from (\ref{completehyperaction}) that if the $q^{au}$ vanish
then the complex structure scalars $z^i$ have no potential and so become flat directions. 
The top two plots of Fig. \ref{fall} give the results of a run showing how the first components of
two different hypermultiplets get driven toward each other as $z$ remains small. This confirms
the expectation that the hyper-multiplets follow the flat directions present at $z=0$, so placing the
gauge theory in the Higgs phase and allowing the conifold transition to complete.
\begin{table}
\center
 \begin{tabular}{||l|c|c|c|c|c|c|c|c|c||} \hline
   Plot & $z$ & $q^{11}$ & $q^{12}$ & $q^{13}$ & $q^{14}$ & $q^{21}$ & $q^{22}$ & $q^{23}$ & $q^{24}$ \\ \hline
   $1^{st}$ & 0.01 & 0.1 & 0.15 & 0.2 & 0.25 & 0.15 & 0.2 & 0.25 & 0.3 \\ \hline
   $2^{nd}$ & 0.95 & 1$^{-4}$ & 1.5x$10^{-4}$ & 2x$10^{-4}$ & 2.5x$10^{-4}$ & 2x$10^{-4}$ 
            & 3x$10^{-4}$ & 4x$10^{-4}$ & 5x$10^{-4}$ \\ \hline
   $3^{rd}$ & 0.65 & 0.01 & 0.015 & 0.02 & 0.025 & 0.02 & 0.03 & 0.04 & 0.05 \\ \hline
 \end{tabular}
\caption{Table showing the initial values for the fields, corresponding to the plots in Figure \ref{fall}}
\label{tall}
\end{table}

In the second plot we see the case where the hyper-multiplets have reached small values before the 
complex structure modulus has reached the conifold point. 
In this case the potential generated by the $q^{au}$ is too small to 
drive the $z$ to zero and the field theory never reaches the conifold, 
leaving it in the Coulomb phase. This crucially depends on the Hubble damping
being efficient enough to stop the $q^{au}$ from oscillating.
The third plot shows results corresponding to a case where both the complex structure and the
hyper-multiplets oscillate around zero. From the string theory picture, this scenario describes
the CY being stuck at the conifold point, with the oscillations of $z$ keeping $q$ small and vice versa. 
This means that both the three-cycles of the original
CY and the two-cycles of the topologically related CY remain small.

\begin{figure}
\center
\epsfig{file=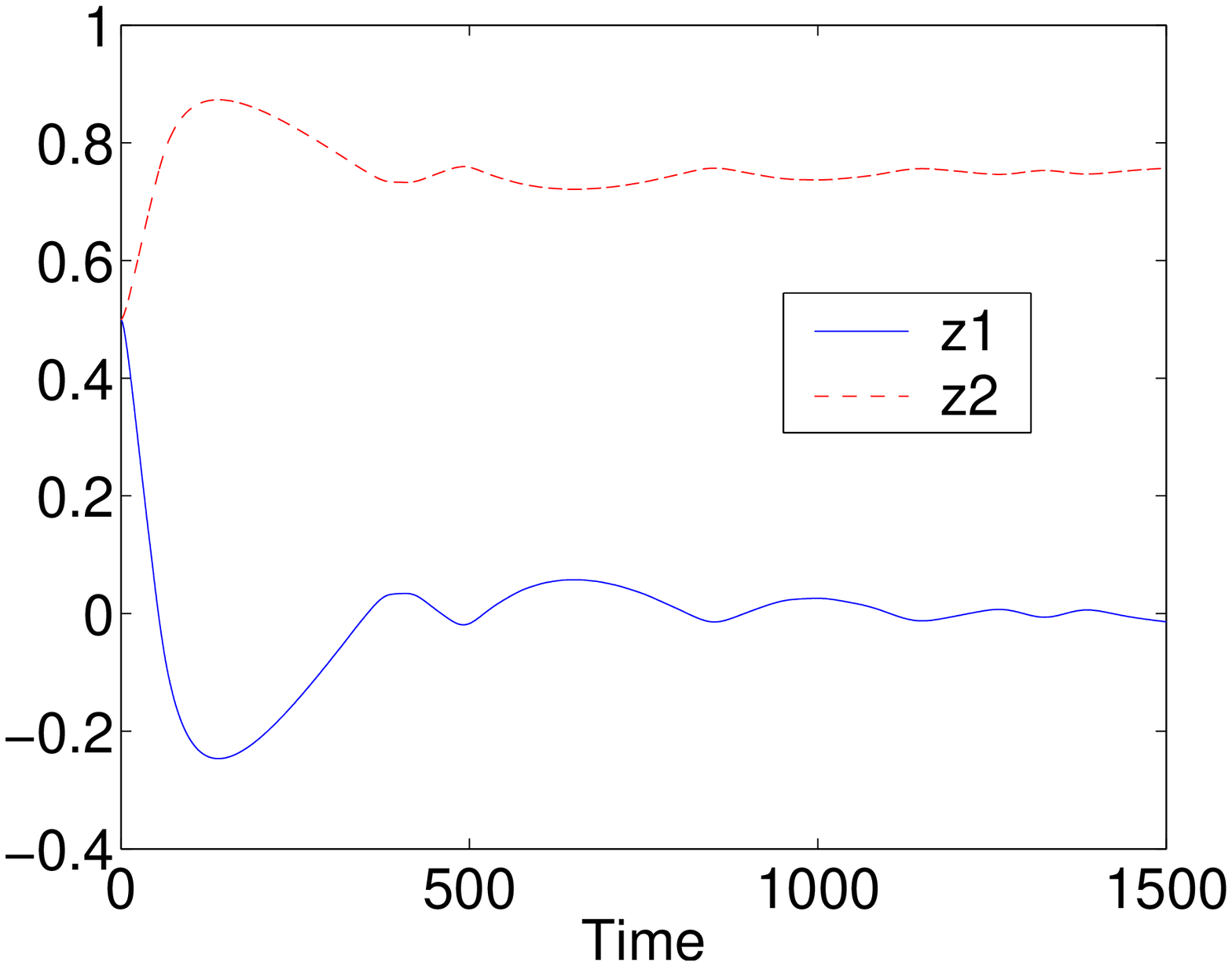,width=7.5cm}
\epsfig{file=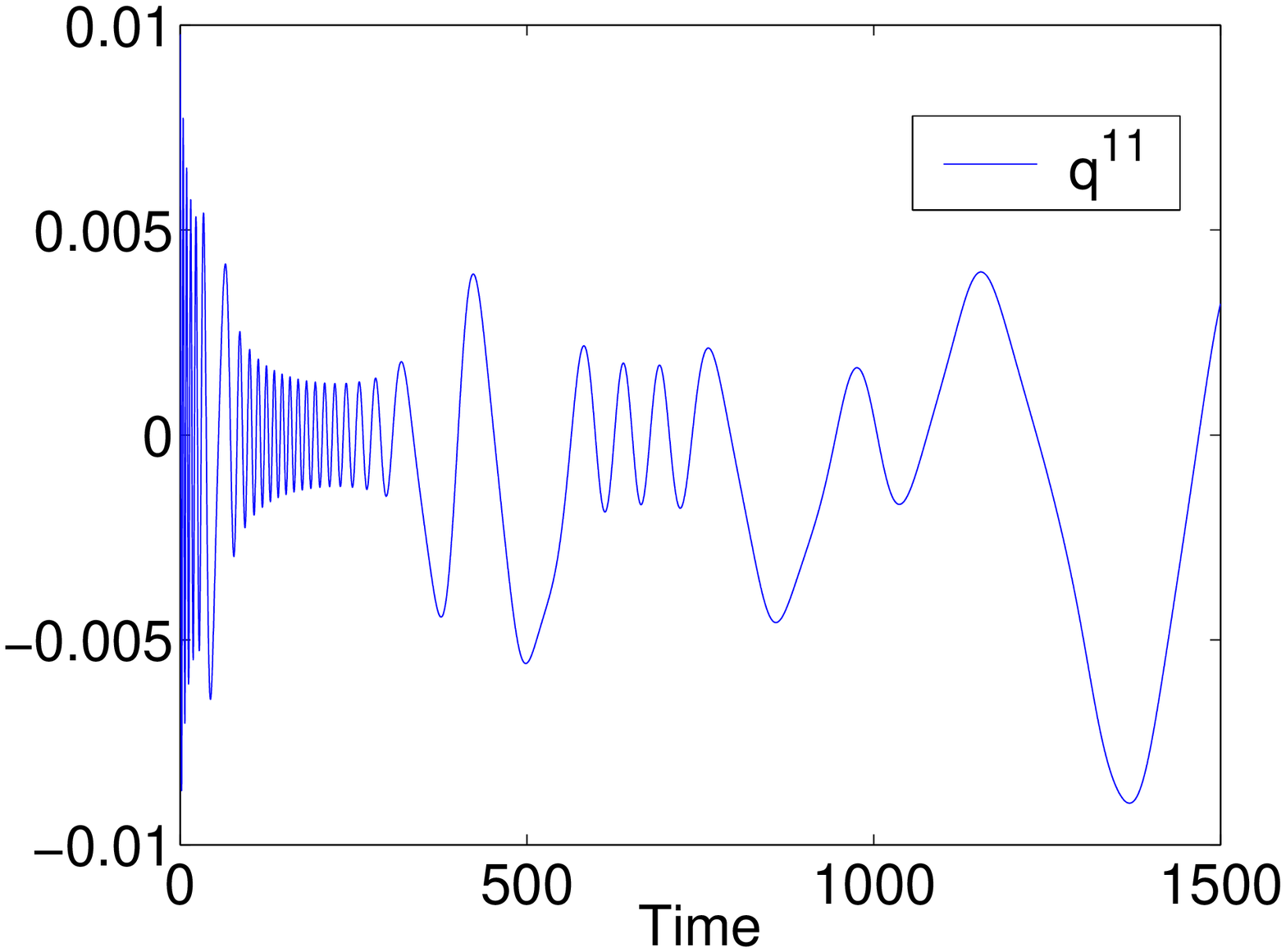,width=7.5cm}
\flushleft
\caption{ Plots showing evolution of $q^{11}$, the wrapped modulus $Re(z^1)$ and unwrapped modulus $Re(z^2)$ with 
initial values given in Table \ref{tcouple}. }
\label{fcouple}
\end{figure}

As discussed earlier it is possible to couple the gauge field super-partners of the wrapped moduli 
to the gauge fields associated with unwrapped moduli through off diagonal elements in the coupling matrix,
and we argued that this created dynamics for those cycles which were not even being wrapped.
To give an example of this
consider an unwrapped modulus $z^2$ along with the original $z=z^1$, but now we move away from minimal coupling,
taking the coupling matrix to be of the form 
\be
N_{IJ}=\left( \begin{array}{ccc} 1 &0 &0 \\ 0 &-1  &-\half \\ 0 &-\half &-1  \end{array}  \right).
\ee
Fig. \ref{fcouple} shows the evolution of both the complex structure moduli $z^1$, $z^2$,
along with the representative component $q^{11}$
for the initial conditions given in Table \ref{tcouple}. We see that, as expected, the evolution
of the wrapped cycle corresponding to $z^1$ has created a potential for the unwrapped cycle
given by $z^2$. In this particular case the unwrapped cycle has also settled down
to a particular size with its final value clearly depending on the choice of coupling
matrix, and hence on the CY, as well as on the initial conditions. For this particular
example the final state sees the hypermultiplets remaining small along with the value of
$z^1$, the wrapped cycle. That is to say, the fields remain at the conifold point.

\begin{table}
\center
 \begin{tabular}{||l|c|c|c|c|c|c|c|c|c|c||} \hline
   Field & $z^1$ & $z^2$ & $q^{11}$ & $q^{12}$ & $q^{13}$ & $q^{14}$ & $q^{21}$ & $q^{22}$ & $q^{23}$ & $q^{24}$ \\ \hline
   Value & 0.5 & 0.5 & 0.01 & 0.015 & 0.02 & 0.025 & 0.015 & 0.02 & 0.025 & 0.03 \\ \hline
 \end{tabular}
\caption{Table showing the initial values of the fields plotted in Figure \ref{fcouple}.}
\label{tcouple}
\end{table}

\section{Conclusion}
\label{sec:conclusions}
In this paper we have examined the effect of a particular topological
transition of the internal manifold in a Calabi-Yau compactification
of IIB string theory.  The transition in question is the conifold
transition where one Calabi-Yau gets transformed to another via the
shrinking and expansing of two-cycles and three-cycles. The question
we try to answer here is how do these cycles evolve in a cosmological
context? And is it possible to fix the moduli associated with these
cycles?

By constructing the effective action for IIB supergravity with
D3-branes wrapping the three-cycles we have seen that the brane
wrapping states can be described by charged hyper-multiplets,
$q^{au}$, in four dimensions \cite{Strominger:1995cz} with the charges
corresponding to the cycles which are being wrapped.  As was to be
expected, exciting the wrapping states drove the three-cycles toward
zero volume by creating a potential for associated complex structure
moduli, $z^i$.  However, once the three-cycles are at zero volume then
the $q^{au}$ have flat directions so could pick up a vev, placing the
gauge theory in the Higgs phase and taking the Calabi-Yau succesfully
through the conifold transition. It was also possible that the wrapped
D3-branes/anti-branes could annihilate  before the three-cycles are
able to reach the conifold point, in which case there is no longer a
force driving the cycles to collapse and the theory remains in the
Coulomb phase.  A third possibility is that both the three-cycles and
two-cycles are driven to zero dynamically, so that the Calabi-Yau ends
up near the conifold point.  By numerically evolving the equations of
motion for the effective action we have seen that all three cases are
possible, depending on initial conditions of the fields. 
In the absence of information about the processes causing the brane
wrapping to occur it is impossible to state which outcome is more likely.
However, it seems
that in the case where the fields are of the same order of magnitude
the generic evolution consists of the fields oscillating around
zero. In this sense we would expect the internal space to typically be
near the transition in a cosmological context. Taking the point of
view given in \cite{Kofman:2004yc} that quantum fields are always
excited one would again conclude that it is more likely that the
fields should remain near the conifold point.  We note also that this
picture makes our approximation of small $q^{au}$ consistent and so we
expect the quaternion-K\"ahler metric $h_{uv}$ to be well described by
the flat metric we used. This raises another issue, namely what are the
typical values that these scalar fields should take? This is clearly
an important point as the values of these fields can dramatically
change the evolution of the system.

We have also seen how the structure of the Calabi-Yau, through its
coupling matrix, can create a potential even for those cycles which are
not being wrapped. Depending on the form of this matrix then, it is
possible for these other complex structure moduli to be involved in
the evolution.

As we have performed the simulations in a cosmological context, the
oscillations of the fields have been damped by Hubble friction. This
is in fact crucial if, for example, one wants to complete the conifold
transition and end up in the Higgs phase. Without Hubble damping both
the complex structure and the hyper-multiplets oscillate about zero in
an apparently chaotic manner, thereby never reaching the state
$z^i=0$, $q^{au}\neq 0$, which would be a completed conifold
transition. Hence, dynamically we would not expect a conifold transition
to occur in Minkowski space.

In our effective theory we have made a rather arbitrary choice for
pre-potential ${\cal F}$ and not made any attempt to connect it to an
existing Calabi-Yau.  It would be a worthwhile exercise to make a more
precise link with string theory and use results for a particular
Calabi-Yau to tell us which ${\cal F}$ to use.  It would also be
useful to have a clearer understanding of the role the coupling matrix
plays in generating dynamics for those cycles which are not wrapped.
Something which we have not touched upon is that of causality and the
Kibble mechanism \cite{Kibble:1976sj}. In the real Universe we only
expect homogeneity on sufficiently large scales while on smaller scales we
would expect the fields to take on different value as
required by causality \cite{Kibble:1976sj}. What this means for the
conifold transition is that after its completion the broken U(1) gauge
symmetry will generate cosmic strings. However, as emphasized in
\cite{Achucarro:1998er} these strings are rather different to the
usual Nielsen-Olesen vortices \cite{Nielsen:1973cs} in that they are
unstable to expansion of their core.  It would therefore be
interesting to study the dynamics of these strings to see how the
instability-timescale alters the formation of string networks.

\vspace{1cm}
\noindent
{\large\bf Note added} While this paper was being prepared some related work
was presented on the archive which also
analyzes the cosmology of conifold transitions, albeit in a different context
\cite{Mohaupt:2004pq}\cite{Mohaupt:2004pr}. These papers study 
the five-dimensional cosmology following from M-theory on a Calabi-Yau space
as it goes through a conifold transition, including the effect
of M2-branes wrapped on two-cycles.

\vspace{1cm}
\noindent
{\large\bf Acknowledgements} A.~L.~and P.~M.~S.~are supported by PPARC
Advanced Fellowships. E.~P.~is supported by a PPARC studentship.

\vskip 1cm
\appendix{\noindent\Large \bf Appendices}
\renewcommand{\theequation}{\Alph{section}.\arabic{equation}}
\setcounter{equation}{0}

\section{Conventions}
\label{AppA}
In this paper we take the metric to have signature ( - + + + ...). We define the Levi-Civita tensor $\epsilon$ and the volume
form $\eta$ as 
\ba
\epsilon_{012..}=+1\\
dx^{\mu}\wedge dx^{\nu} \wedge ...=-\epsilon^{\mu\nu...}\eta\\
\eta=dx^0 \wedge dx^1 \wedge ...
\ea
The Hodge dual is defined as
\ba
\star F_{p} = \frac{1}{p!(d-p)!} F_{\mu_1\mu_2...\mu_p}\epsilon^{\mu_1\mu_2...\mu_p}_{\qquad\qquad\mu_{p+1}\mu_{p+2}...\mu_d}
                                                 dx^{\mu_{p+1}}\wedge dx^{\mu_{p+2}}\wedge ... dx^{\mu_d}.
\ea
Giving for p-forms $\alpha$ and $\beta$
\ba
\alpha\wedge\star\beta&=&(\alpha\lrcorner\beta)\eta=\frac{1}{p!}\alpha^{\mu\nu ...}\beta_{\mu\nu ...}\eta
\ea


\end{document}